\documentclass[aps,pre,floats,twocolumn,showpacs,superscriptaddress]{revtex4}

\usepackage{cancel}
\usepackage{graphicx,epsfig}
\usepackage{times}
\usepackage{graphics,dcolumn,bm,fleqn,epic,eepic,float}
\usepackage{amssymb,amsmath,multirow,rotate,color}
\usepackage{subfigure}
\begin{document}

\title{Traffic-driven Epidemic Spreading in Finite-size Scale-Free Networks}
\author{S. Meloni}
\affiliation{Department of Informatics and Automation, University of Rome "Roma Tre", Via della Vasca Navale, 79
00146, Rome, Italy}

\author{A. Arenas}
\affiliation{Departament d'Enginyeria Inform\`atica i Matem\`atiques, Universitat Rovira i Virgili, 43007 Tarragona, Spain}
\affiliation{Institute for Biocomputation and Physics of Complex Systems, University of Zaragoza, 50009 Zaragoza, Spain}    

\author{Y. Moreno}\email{yamir.moreno@gmail.com}
\affiliation{Institute for Biocomputation and Physics of Complex Systems, University of Zaragoza, 50009 Zaragoza, Spain}    
\affiliation{Department of Theoretical Physics, University of Zaragoza, 50009 Zaragoza, Spain}

\date{\today}

\begin{abstract}

The study of complex networks sheds light on the relation between the structure and function of complex systems. One remarkable result is the absence of an epidemic threshold in infinite-size scale-free networks, which implies that any infection will perpetually propagate regardless of the spreading rate. The vast majority of current theoretical approaches assumes that infections are transmitted as a reaction process from nodes to all neighbors. Here we adopt a different perspective and show that the epidemic incidence is shaped by traffic flow conditions. Specifically, we consider the scenario in which epidemic pathways are defined and driven by flows. Through extensive numerical simulations and theoretical predictions, it is shown that the value of the epidemic threshold in scale-free networks depends directly on flow conditions, in particular on the first and second moments of the betweenness distribution given a routing protocol. We consider the scenarios in which the delivery capability of the nodes is bounded or unbounded. In both cases, the threshold values depend on the traffic and decrease as flow increases. Bounded delivery provokes the emergence of congestion, slowing down the spreading of the disease and setting a limit for the epidemic incidence.
Our results provide a general conceptual framework to understand spreading processes on complex networks.

\end{abstract}


\maketitle

\section{Introduction}

Scale-free networks \cite{blmch06,dgm08, ba99} are characterized by the presence of hubs, which are responsible for several striking properties for the propagation of information, rumors or infections \cite{psv01,llm01,mpsv01,n02, bbpsv04,glmp08}. Theoretical modeling of how diseases spread in complex networks is largely based on the assumption that the propagation is driven by reaction processes, in the sense that the transmission occurs from every infected through all its neighbors at each time step, producing a diffusion of the epidemics on the network. However, this approach overlooks the notion that the network substrate is a fixed snapshot of all the possible connections between nodes, which does not imply that all nodes are concurrently active. Many networks observed in nature \cite{blmch06,dgm08}, including those in society, biology and technology, have nodes that temporally interact only with a subset of its neighbors \cite{asbs00,nfb02}. For instance, hub proteins do not always interact with all their neighbor proteins at the same time \cite{han_etal09}, just as individuals in a social network \cite{stanley} do not interact simultaneously with all of their acquaintances. Likewise, Internet connections being utilized at a given time depends on the specific traffic and routing protocols. Given that transport is one of the most common functions of networked systems, a proper consideration of this issue will irreparably affect how a given dynamical process evolves. 

Our knowledge of the mechanisms involved in disease propagation has improved in the last several years \cite{egamstw04,cpv07,cv08}. Recent works have to some extend surmounted the problem of link concurrency through agent-based modeling approaches \cite{egamstw04} or by explicitly implementing data-driven simulation tools \cite{cbbv07}. These very detailed computer models have not been yet analyzed in a general theoretical framework. Here we introduce a theoretical approach to investigate the outcome of an epidemic spreading process driven by transport instead of reaction events. To this end, we analyze a paradigmatic abstraction of epidemic contagion, the so-called Susceptible-Infected-Susceptible (SIS) model \cite{m07} (see {\it Methods}), which assumes that contagion occurs through the eventual contact or transmission between connected partners that are using their connections at the time of propagation. This is achieved by considering a quantized interaction at each time step. Mathematically, we set up the model in a flow scenario where contagion is carried by interaction packets traveling across the network.

We consider two possible scenarios that encompass most of real traffic situations: i) unbounded delivery rate, and ii) bounded delivery rate, of packets per unit time. We derive the equation governing the critical threshold for epidemic spreading in SF networks, which embeds, as a particular case, previous theoretical findings. For unbounded delivery rate, it is shown that the epidemic threshold decreases in {\it finite} SF networks when traffic flow increases. In the bounded case, nodes accumulate packets at their queues when traffic flow overcomes the maximal delivery rate, i.e. when congestion arises. From this moment on, the results show that both the epidemic threshold and the infection prevalence are bounded due to congestion.

\section{Results and Discussion}

We first generate two different types of SF networks. On one hand, we construct random SF networks where no correlations are present using the configuration model \cite{blmch06,dgm08}. On the other hand, we also generate small-world, SF and highly clustered networks $-$ all properties found in many real-world networks \cite{blmch06,dgm08} such as the Internet $-$ using a class of recently developed network models \cite{skb08,bkc09}, in which nearby nodes in a hidden metric space are connected. This metric space can represent social, geographical or any other relevant distance between the nodes of the simulated networks. Specifically, in the model we use, nodes are uniformly distributed in a one-dimensional circle and assigned an expected degree $k$ from a power law distribution $P(k)\sim k^{-\gamma}$. Pairs of nodes separated by a distance $d$ are linked with a probability $r(d;k;k')=(1+d/d_c)^{-\alpha}$, where $\alpha>1$ and $d_c\sim kk'$ (see {\it Methods} for further details).  In most of our simulations, we fixed $\langle k \rangle=3$ and $\alpha=2$. Once the networks are built up, the traffic process is implemented in the following way. At each time step, $p=\lambda N$ new packets are created with randomly chosen origins and destinations. The routing of information is modeled through even a shortest path delivery strategy or a greedy algorithm \cite{bkc09, bk09}. In the latter, we make use of the second class of SF networks and a node $i$ forwards a packet to node $j$ in its neighborhood, which is the closest node (in the hidden metric space) to the final packet destination.

To account for link concurrency, we consider that two nodes do not interact at all times $t$, but only when they exchange at least a packet. Therefore, we assume that the epidemic can spread between nodes every time an interaction takes place. This situation is reminiscent of disease transmission on air transportation networks; if an infected individual did not travel between two cities, then regardless of whether or not those cities are connected by a direct flight, the epidemic will not spread from one place to the other. In this way, although a node can potentially interact with as many contacts as it has and as many times as packets it exchanges with its neighbors, the effective interactions are driven by a second dynamics (traffic). The more packets travel through a link, the more likely the disease will spread through it. On the other hand, once an interaction is at work, the epidemics spreads from infected to susceptible nodes with probability $\beta$. For example, if at time $t$ node $i$ is infected and a packet is traveling from node $i$ to one of its neighbors node $j$, then at the next time step, node $j$ will be infected with probability $\beta$. Therefore, susceptible and infected states are associated with the nodes, whereas the transport of packets is the mechanism responsible for the propagation of the disease at each time step.

\subsection{Unbounded Delivery Rate}

\begin{figure*}
\begin{center}
\epsfig{file=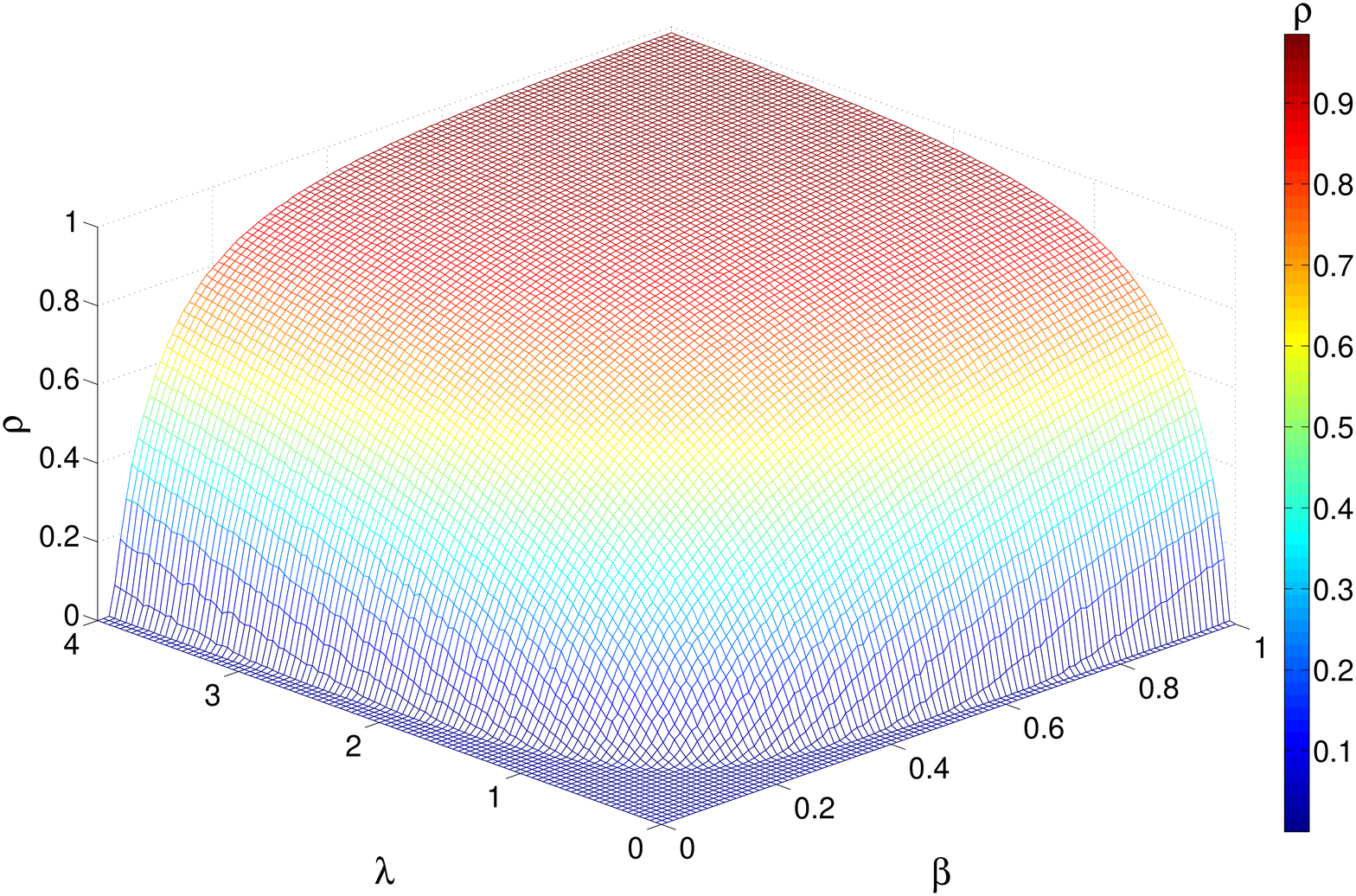, width=8cm, angle=0}
\end{center}
\caption{Dependence of epidemic incidence on traffic conditions for unbounded delivery rate. The density of infected nodes, $\rho$, is shown as a function of the spreading rate $\beta$ and the intensity of flow $\lambda$ in SF networks. Flow conditions (controlled by $\lambda$) determine both the prevalence level and the values of the epidemic thresholds. Increasing the number of packets traveling through the system has a malicious effect: the epidemic threshold decreases as the flow increases. Each curve is an average of 100 simulations starting from an initial density of infected nodes $\rho_0=0.05$. The results corresponds to the greedy routing scheme and the network is made up of $10^3$ nodes using the model in \cite{bkc09}. The remaining parameters are $\alpha=2$, $\gamma=2.6$ and $\langle k \rangle=3$.}
\label{fig1}
\end{figure*}

In this situation, congestion can not arise in the system. Fig.\ \ref{fig1} shows the results for the stationary density of infected nodes $\rho$ as a function of $\beta$ and the traffic generation rate $\lambda$ for SF networks. 

The traffic level determines the value of both the epidemic incidence and the critical thresholds. We observe the emergence of an epidemic threshold under low traffic conditions. This implies that for a fixed value of $\lambda$, the epidemic dies out if the spreading rate is below a certain critical value $\beta_c(\lambda)$. More intense packet flows yield lower epidemic thresholds. The reason for the dependence of the critical spreading rates on $\lambda$ is rooted in the effective topological paths induced by the flow of packets through the network. At low values of $\lambda$, there are only a few packets traveling throughout the system, so the epidemic simply dies out because many nodes do not participate in the interaction via packets exchanges. As $\lambda$ grows, more paths appear between communicating nodes, thus spreading the infection to a larger portion of the network. Therefore, in traffic-driven epidemic processes the infection is constrained to propagate only through links that transmit a packet, and thus the number of attempts to transmit the infection depends on the flow conditions at a local level, namely, on the number of active communication channels at each time step. As a consequence, the effective network that spreads the infection is no longer equivalent to the complete underlying topology.  Instead, it is a map of the dynamical process associated with packet traffic flow. The conclusion is that the disease propagation process has two dynamical components: one intrinsic to the disease itself ($\beta$) and the other to the underlying traffic dynamics (the flow).

To theorize about these effects we next formulate the analytical expression for the dependence of the epidemic threshold on the amount of traffic injected into the system, following a mean-field approach akin to the conventional analysis of the reaction driven case. Mathematically, the fraction of paths traversing a node given a certain routing protocol \cite{gdvca02}, the so-called algorithmic betweenness, $b^{k}_{\text{alg}}$, defines the flow pathways. Let us consider the evolution of the relative density, $\rho_k(t)$, of infected nodes with degree $k$. Following the heterogeneous mean-field approximation \cite{psv01}, the dynamical rate equations for the SIS model are 
\begin{equation}
  \partial_t \rho_k(t) = -\mu \rho_k(t) +\beta \lambda b^{k}_{\text{alg}} N \left[
  1-\rho_k(t) \right] \Theta(t). 
  \label{mf_eqs}
\end{equation}
The first term in Eq.\ (\ref{mf_eqs}) is the recovery rate of infected individuals (we set henceforth $\mu=1$). The second term takes into account the probability that a node with $k$ links
belongs to the susceptible class, $[1-\rho_k(t)]$, and gets the infection via packets traveling from infected nodes. The latter process is proportional to the spreading probability $\beta$, the probability $\Theta(t)$ that a packet travels through a link pointing to an
infected node and the number of \emph{packets} received by a node of degree $k$. This, in turns, is proportional to the total number of packets in the system, $\sim\lambda N$, and the algorithmic betweenness of the node,  $b^{k}_{\text{alg}}$. Note that the difference with the standard epidemic spreading model is given by these factors, as now the number of contacts per unit time of a node is not proportional to its connectivity but to the number of packets that travel through it. Finally, $\Theta(t)$ takes the form
\begin{equation}
 \Theta(t)=\frac{\sum_k  b^{k}_{\text{alg}} P(k) \rho_k(t)}{\sum_k  b^{k}_{\text{alg}} P(k)}.
  \label{eq2a}
\end{equation}
Eq. (\ref{mf_eqs}) has been obtained assuming: (i) that the network is uncorrelated $P(k' |k)=k'P(k')/\langle k \rangle$, and (ii) that the algorithmic flow between the classes of nodes of degree $k$ and $k'$ factorizes $b^{kk'}_{\text{alg}} \sim b^{k}_{\text{alg}}b^{k'}_{\text{alg}}$. Although no uncorrelated networks exist, this approximation allows us to identify the governing parameters of the proposed dynamics. The second approximation is an upper bound to the actual value of the $b^{kk'}_{\text{alg}}$, whose mathematical expression is, in general, unknown. The validity of the theory even with these approximations is notable as confirmed by the numerical simulations.

\begin{figure*}
\begin{center}
\epsfig{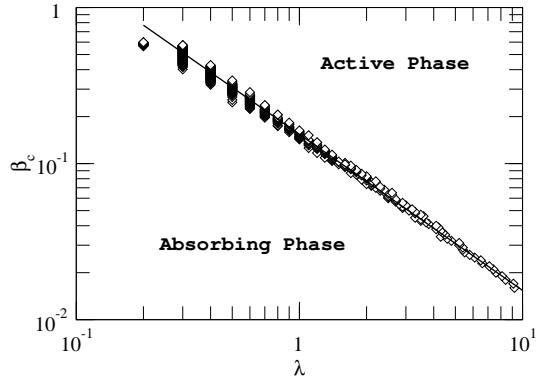}
\end{center}
\caption{Comparison between numerical and theoretical critical points. Log-log plot of the critical thresholds, $\beta_c$, as a function of the rate at which packets are injected into the system, $\lambda$. Two regions are differentiated: an active and an absorbing phase as indicated. The solid line corresponds to Eq.\ (\ref{threshold}) with  $\frac{\langle b_{\text{alg}} \rangle}{\langle b^2_{\text{alg}} \rangle}\frac{1}{N}=0.154$. The agreement is remarkable even though Eq.\ (\ref{threshold}) is derived using a mean field approach. The underlying network, infection spreading mechanism and routing protocol are the same as in Fig.\ \ref{fig1}. Each curve is an average of $10^2$ simulations. Remaining parameters are the same as in Fig.\ \ref{fig1}.} 
\label{fig2}
\end{figure*}

By imposing stationarity 
[$\partial_t \rho_k(t) =0$], Eq.\ (\ref{mf_eqs}) yields
\begin{equation}
  \rho_k=\frac{\beta \lambda b^{k}_{\text{alg}} N\Theta}{1+\beta \lambda b^{k}_{\text{alg}} N\Theta},
  \label{eq2b}
\end{equation}
from which a self-consistent equation for $\Theta$ is obtained as
\begin{equation}
 \Theta=\frac{1}{{\sum_k  b^{k}_{\text{alg}} P(k)}}\sum_k  \frac{(b^{k}_{\text{alg}})^2 P(k) \beta \lambda N \Theta}{1+\beta \lambda b^{k}_{\text{alg}} N\Theta}.
  \label{eq2c}
\end{equation}
The value $ \Theta = 0$ is always a solution. In order to have a non-zero
solution, the condition 
\begin{equation}
  \frac{1}{{\sum_k  b^{k}_{\text{alg}} P(k)}}  \frac{d}{d \Theta} \left. \left( \sum_k  \frac{(b^{k}_{\text{alg}})^2 P(k) \beta \lambda N \Theta}{1+\beta \lambda b^{k}_{\text{alg}} N\Theta} \right) \right|_{\Theta=0} > 1
\end{equation}
must be fulfilled, from which the epidemic threshold is obtained as
\begin{equation}
  \beta_c = \frac{\langle b_{\text{alg}} \rangle}{\langle b^2_{\text{alg}} \rangle}\frac{1}{\lambda N},
  \label{threshold}
\end{equation}
below which the epidemic dies out, and above which there is an endemic state. In Fig.\ \ref{fig2} a comparison between the theoretical prediction and numerical observations is presented. Here, we have explicitly calculated the algorithmic betweenness for the greedy routing as it only coincides with the topological betweenness for shortest paths routing. The obtained curve separates two regions: an absorbing phase in which the epidemic disappears, and an active phase where the infection is endemic.

Equation\ (\ref{threshold}) is notably simple but has profound implications: the epidemic threshold decreases with traffic and eventually vanishes in the limit of very large traffic flow in finite systems, in contrast to the expected result of a finite-size reminiscent threshold in the classical reactive--diffusive framework. Admittedly, this is a new feature with respect to previous results on epidemic spreading in SF networks. It is rooted in the increase of the effective epidemic spreading rate due to the flow of packets. This is a genuine effect of traffic-driven epidemic processes and generalizes the hypothesis put forward in the framework of a reaction-diffusion process \cite{cpv07} on SF networks. It implies that an epidemic will pervade the (finite) network whatever the spreading rate is if the load on it is high enough. Moreover, Eq.\ (\ref{threshold}) reveals a new dependence. The critical threshold depends on the topological features of the graph, but at variance with the standard case, through the first two moments of the algorithmic betweenness distribution. As noted above, the algorithmic betweenness of a node is given by the number of packets traversing that node given a routing protocol. In other words, it has two components: a topological one which is given by the degree of the node and a dynamical component defined by the routing protocol.

Within our formulation, the classical result \cite{psv01}
\begin{equation}
\beta_c=\frac{\langle k \rangle}{\langle k^2 \rangle},
\label{result_psv}
\end{equation}
can be obtained for a particular protocol and traffic conditions, although we note that the microscopic dynamics of our model is different from the classical SIS. To see this, assume a random protocol. If packets of information are represented as $w$ random walkers traveling in a network with average degree $\langle k \rangle$, then under the assumption that the packets are not interacting, it follows that the average number of walkers at a node $i$ in the stationary regime (the algorithmic betweenness) is given by \cite{nr04,mglm08} $b^i_{\text{alg}}=\frac{k_i}{N\langle k \rangle}w$. The effective critical value is then $(\beta  \lambda)_c=<k>^2/(<k^2>w)$, that recovers, when $\omega=\langle k \rangle$, the result in Eq.\ (\ref{result_psv}).

\begin{figure*}
\begin{center}
\epsfig{file=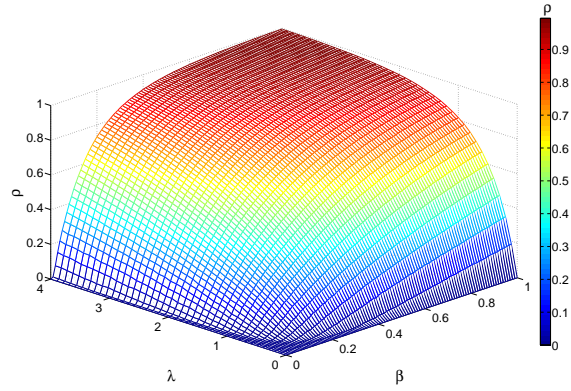, width=8cm, angle=0}
\end{center}
\caption{Density of infected nodes, $\rho$, as a function of traffic flow (determined by $\lambda$) and the epidemic spreading rate $\beta$ for random SF networks and a shortest paths routing scheme for packets delivery. Each point is the result of 100 averages over different networks and initial conditions. The network has a degree distribution with an exponent $\gamma=2.7$.} 
\label{fig3}
\end{figure*}

Our results are robust for other network models and different routing algorithms. We have also made numerical simulations of the traffic-driven epidemic process on top of Barab\'asi-Albert and random SF networks implementing a shortest paths delivery scheme. In this case, packets are diverted following the shortest path (in the actual topological space) from the packets' origins to their destinations. The rest of model parameters and rules for epidemic spreading remain the same. Figure\ \ref{fig3} shows the results obtained for random SF networks. As can be seen, the phenomenology is the same: the epidemic threshold depends on the amount of traffic in the network such that the higher the flow is, the smaller the epidemic threshold separating the absorbing and active phases. On the other hand, for processes in which the delivery of packets follows a shortest path algorithm, Eq. (\ref{threshold}) looks like
\begin{equation}
  \beta_c = \frac{\langle b_{\text{top}} \rangle}{\langle b^2_{\text{top}} \rangle}\frac{1}{\lambda N},
  \label{threshold_top}
\end{equation}
where $b_{\text{top}}$ is the topological betweenness.

\subsection{Bounded Delivery Rate}

Equation (\ref{threshold_top}) allows us to investigate also the equivalent scenario in the presence of congestion. 
Let us consider the same traffic process above but with nodes having queues that can store as many 
packets as needed but can deliver, on average, only a finite number of them at each time step. It is known that there is a critical value of $\lambda$ above which the system starts to congest\cite{gdvca02}
\begin{equation}
\lambda_c = \frac{(N-1)}{b_{\text{alg}}^{*}}.
\label{lam}
\end{equation}
Equation (\ref{lam}) gives the traffic threshold that defines the onset of congestion, which is governed by the node with maximum algorithmic betweenness $b_{\text{alg}}^{*}$.  Substituting (\ref{lam}) in (\ref{threshold}) we obtain a critical threshold for an epidemic spreading process bounded by congestion. Increasing the traffic above $\lambda_c$ will gradually congest all the nodes in the network up to a limit in which the traffic is stationary and the lengths of queues grow without limit. 

To illustrate this point, let us assume that the capacities for processing and delivering information are heterogeneously distributed \cite{zlpy05,sclts07,rmit08} so that the larger the number of paths traversing a node, the larger its capability to deliver the packets. Specifically, each node $i$ of the network delivers at each time step a maximum of $\lceil c_i=1+ k_i^{\eta}\rceil$ packets, where $\eta$ is a parameter of the model. In this case, the critical value of $\lambda$ in Eq.(\ref{lam}) is multiplied by the maximum delivery capacity \cite{zlpy05}. Moreover, without loss of generality, we will explore the behavior of the model in random SF networks where the routing is implemented by shortest paths $b_{\text{alg}}=b_{\text{top}}\sim k^{\nu}$, being $\nu$ usually between 1.1 and 1.3 \cite{vespbook08}. The previous assumption for the delivery capability thus allows to explore as a function of $\eta$ the situations in which the delivery rate is smaller or larger than the arrival rate (defined by the algorithmic betweenness). Phenomenologically, these two scenarios correspond to the cases in which the traffic is in a free flow regime (if $\eta>\nu$) or when the network will congest (if $\eta<\nu$). We also note that the adopted approach is equivalent to assume a finite length for the queues at the nodes. 

\begin{figure*}
\begin{center}
\epsfig{file=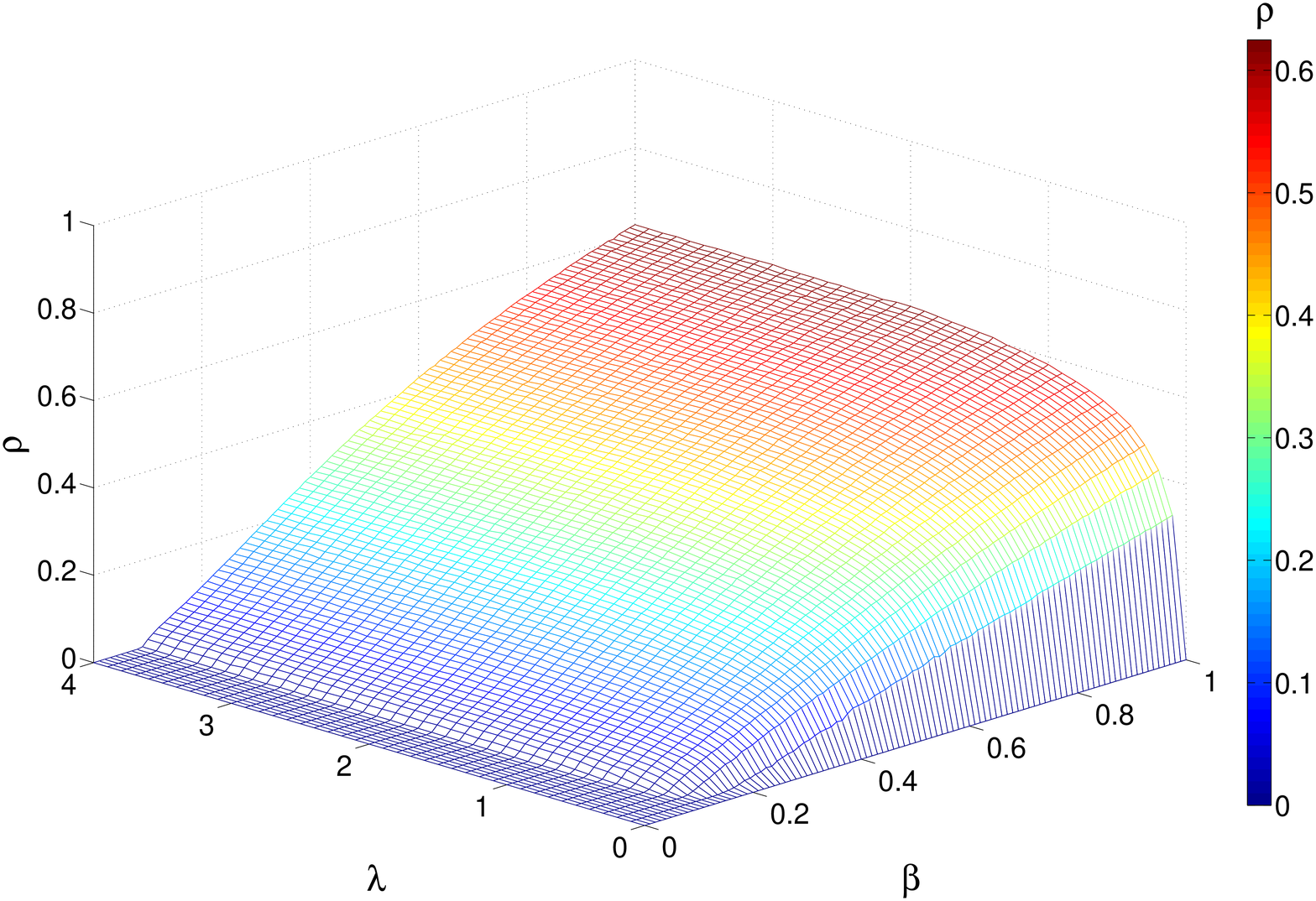, width=8cm, angle=0}
\epsfig{file=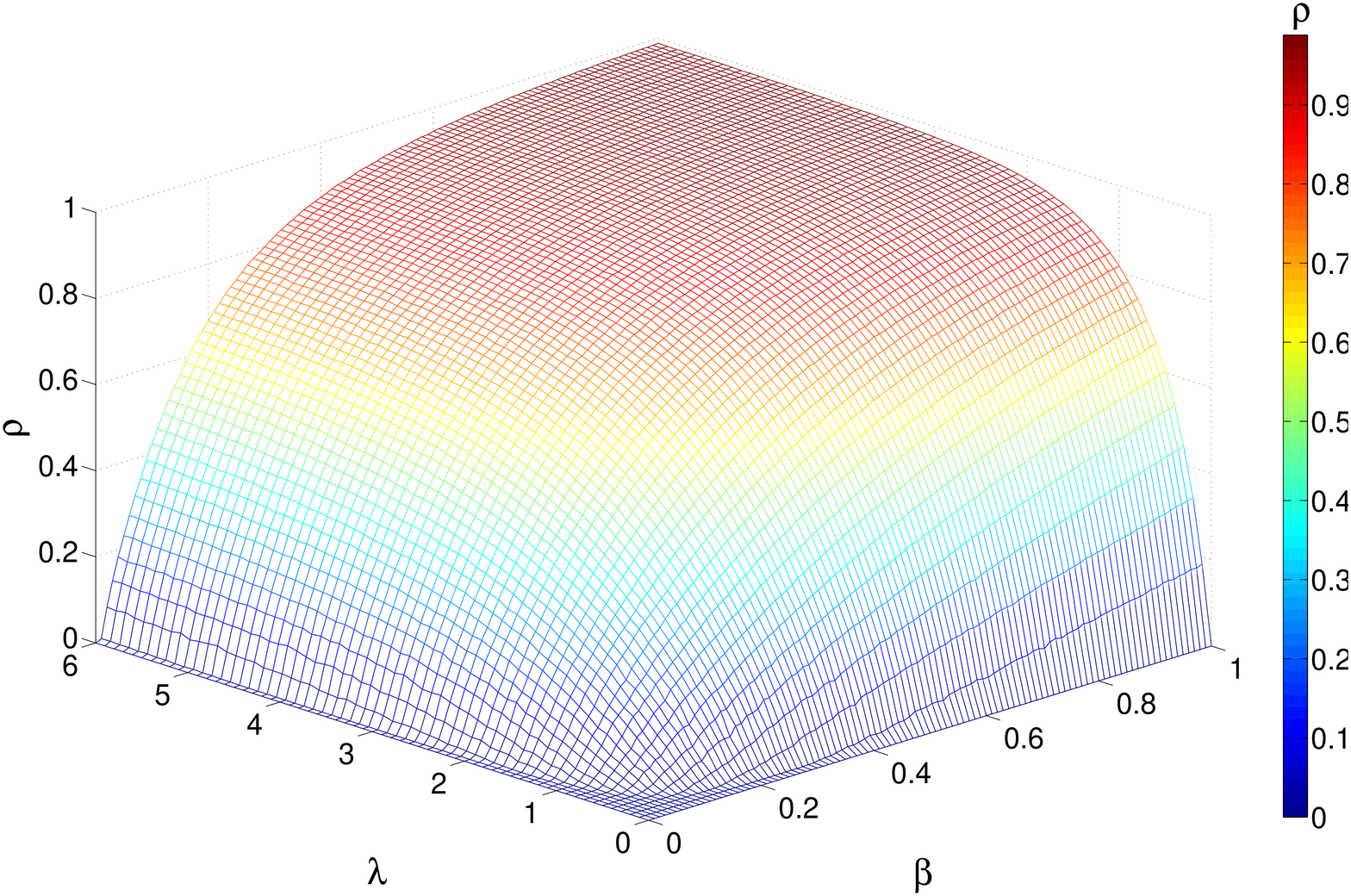, width=8cm, angle=0}
\end{center}
\caption{Dependence of epidemic incidence on traffic conditions for the bounded delivery rate scenario. The figures show the density of infected nodes, $\rho$, as a function of the spreading rate $\beta$ and the intensity of flow $\lambda$ in random SF networks. The top panel corresponds to the case $\eta=0.8$, for which the epidemic threshold is determined by congestion. When the whole network gets congested, no matter whether the value of $\lambda$ is further increased, $\beta_c$ remains constant. In addition, the epidemic incidence saturates. The bottom panel shows the results for $\eta=1.7$. In this case, the epidemic threshold vanishes when $\lambda$ grows and Eq.\ (\ref{threshold}) holds, thus resembling the case of unbounded delivery rate (Fig.\ \ref{fig1}). Each curve is an average of 100 simulations starting from an initial density of infected nodes $\rho_0=0.05$. The network is a random SF network made up of $10^3$ with $\gamma=2.7$ and $\langle k \rangle=3$.}
\label{fig4}
\end{figure*}

Figure \ref{fig4} shows the results for the evolution of the average density of infected nodes, $\rho$, as a function of the spreading rate $\beta$ and the rate at which packets are generated $\lambda$ for two different values of $\eta$ using a shortest path delivery scheme on top of random SF networks. For $\eta=0.8$, the epidemic incidence is significantly small for all values of the parameters $\lambda$ and $\beta$ as compared with the results obtained when the rate of packets delivery is unbounded. On the contrary, when $\eta=1.7$ the phase diagram is qualitatively the same as for the unbounded case, including the result that the epidemic incidence vanishes when $\lambda$ is large enough. A closer look at the dynamical evolution unveils an interesting, previously unreported, feature $-$ when the rate at which packets are delivered is smaller than the rate at which they arrive, the average value of infected nodes saturates beyond a certain value of the traffic flow rate $\lambda$. This effect is due to the emergence of traffic congestion. When the flow of packets into the system is such that nodes are not able to deliver at least as many packets as they receive, their queues start growing and packets pile up. This in turns implies that the spreading of the disease becomes less efficient, or in other words, the spreading process slows down. The consequence is that no matter whether more packets are injected into the system, the average level of packets able to move from nodes to nodes throughout the network is roughly constant and so is the average level of infected individuals.

\begin{figure*}
\begin{center}
\epsfig{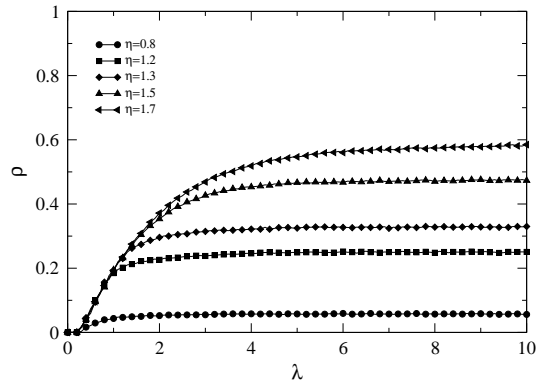}
\end{center}
\caption{Epidemic incidence in traffic-driven epidemic processes with bounded delivery rate. The figure represents the average fraction of infected nodes $\rho$ as a function of $\lambda$ for different delivery rates at fixed $\beta=0.15$. The curves depart from each other when congestion arises and the epidemic incidence saturates soon afterward. The rest of parameters are those of Fig.\ \ref{fig4}.} 
\label{fig5}
\end{figure*}

\begin{figure*}
\begin{center}
\epsfig{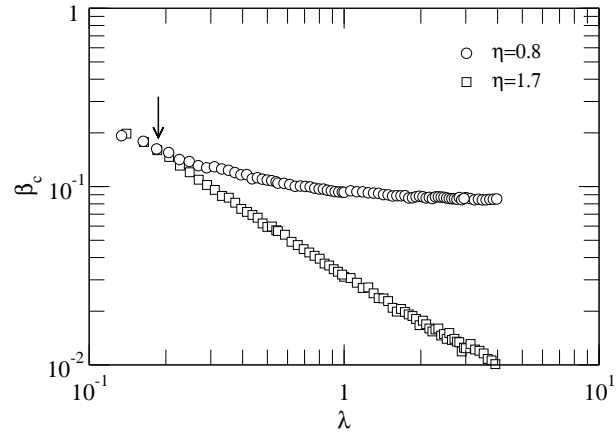}
\end{center}
\caption{Epidemic thresholds as a function of $\lambda$ for two values of $\eta$. The onset of congestion (the arrow in the figure) marks the point, $\lambda_c\approx 0.150$, at which the curve for $\eta=0.8$ departs from Eq.\ (\ref{threshold}), i.e., when the power law dependence breaks down. Soon afterwards congestion extends to the whole network leading to a bounded (from below) epidemic threshold. On the contrary, when the delivery rate is large enough (as in the case of $\eta=1.7$), Eq.\ (\ref{threshold}) holds for all values of $\lambda$, thus resembling the unbounded delivery rate case. Network parameters are those of Fig.\ \ref{fig4}.} 
\label{fig6}
\end{figure*}

Figure\ \ref{fig5} illustrates the phenomenological picture described above. It shows the epidemic incidence $\rho$ for a fixed value of $\beta=0.15$ as a function of $\lambda$ for different values of $\eta$. The figure clearly evidences that congestion is the ultimate reason of the behavior described above. Therefore, the conclusion is that in systems where a traffic process with finite delivery capacity is coupled to the spreading of the disease the {\it epidemic incidence is bounded}. This is good news as most of the spreading processes in real-world networks involves different traffic flow conditions. Further evidence of this phenomenology is given in Fig.\ \ref{fig6}, where we have depicted the epidemic threshold as a function of $\lambda$ for two different values of $\eta$, less and greater than $\nu$. When $\eta<\nu$ congestion arises, and the contrary holds for $\eta>\nu$ where the diagram is equivalent to that of unbounded traffic. The onset of congestion determines the value of $\beta$ above which congestion starts. It is clearly visualized as the point beyond which the power law dependence in Eq.\ (\ref{threshold}) breaks down. The plateau of $\beta_c$ corresponds to the stationary situation of global congestion. 

\section{Conclusions}

In our study, we have computed both analytically and numerically the conditions for the emergence of an epidemic outbreak in SF networks when disease contagion is driven by traffic or interaction flow. We demonstrate that epidemic thresholds are determined by contact flows and that the epidemic incidence is strongly correlated to the emergence of epidemic propagation pathways that are defined and driven by traffic flow. Two general phenomenological scenarios are identified. When the rate at which packets are delivered is large enough or it is not bounded, the epidemic threshold is inversely proportional to the traffic flow in the system. This simple law, which also encompasses a dependency with the topological properties of the graph as given by the algorithmic betweenness has remarkable consequences. One of them is that the epidemic threshold vanishes in the limit of high traffic conditions, even for {\it finite} SF networks and for any routing algorithm. Moreover, within this scenario, the new approach recovers previous results as a particular case of reaction-diffusion spreading. New phenomenological properties appear, in the unified framework, when the system has a finite capability to deliver packets and congestion is possible. In this case, what determines the epidemic incidence is the capability of the network to avoid congestion. If congestion arises, the number of contacts between the system elements decreases, leading to a less efficient spreading of the disease and therefore to a significant reduction of the average number of infected individuals. Additionally, the value of the epidemic threshold depends on the delivery rate $-$ the smaller the delivery rate, the larger the epidemic threshold. Interestingly, our study points out that congestion appears to have a positive effect reducing the critical spreading rate for an epidemic outbreak to occur and its size. 

Finally, we point out that the new approach presented here constitutes an interesting framework to address related problems. For instance, in the context of air-transportation networks \cite{cbbv07,rogerair}, traffic-driven mechanisms of disease propagation could explain the observed differences in the impact of a disease during a year \cite{gekg04}. One might even expect that, due to seasonal fluctuations in flows, the same disease could not provoke a system-wide outbreak if the flow were not high enough during the initial states of the disease contagion. The same reasoning implies that air traffic flow restrictions is an efficient way to content a pandemic spreading. Incorporating the traffic-driven character of the spreading process into current models has profound consequences for the way the system functions. Also the theory could help designing new immunization algorithms or robust protocols; one in particular being quarantining highly sensitive traffic nodes or equivalently control the onset of congestion in networks (control of air traffic flow). Obviously our approximation lacks of the level of detail required to assess more realistic epidemiological models, however it provides a simple framework to anticipate most of the general features to be observed in detailed computational agent-based models.

\section{Methods}

\subsection{Structure}
Scale-free networks are generated by assigning all nodes a random polar angle $\theta$ distributed uniformly in the interval $[0,2\pi)$. The expected degrees of the nodes are then drawn from some distribution $x(k)$ and the network is completed by connecting two nodes with hidden coordinates $(\theta,k)$ and  $(\theta',k')$ with probability $r(\theta,k,\theta',k')=\left(1+\frac{d(\theta,\theta')}{\eta' kk'}\right)^{-\alpha}$, where $\eta'=(\alpha-1)/2\langle k \rangle$, $d(\theta,\theta')$ is the geodesic distance between the two nodes on the circle, and $\langle k \rangle$ is the average degree. Finally, choosing $x(k)=(\gamma-1)k_{0}^{\gamma-1}k^{-\gamma}$, $k>k_0\equiv(\gamma-2)\langle k\rangle/(\gamma-1)$ generates random networks with a power law distribution with exponent $\gamma>2$ .

\subsection{Routing Dynamics}
At each time step, $p=\lambda N$ new packets are created with randomly chosen origins and destinations. For the sake of simplicity, packets are considered non-interacting so that no queues are used. The greedy routing delivers packets such that node $i$, holding a packet whose destination is $l$, dispatches it through node $j$ that is the closest to $l$ in the hidden metric space. Alternatively, one can implement a shortest path algorithm in which packets are delivered through nodes that are closer to their destinations in the network (i.e., not in the hidden metric space). Our results are insensitive to the two routing protocols implemented.

\subsection{Epidemic Dynamics}
We have implemented the Susceptible-Infected-Susceptible model in which each node can be in two possible states: healthy (S) or infected (I). Starting from an initial fraction of infected individuals $\rho_0=I_0/N$, the infection spreads in the system through packet exchanges. A susceptible node has a probability $\beta$ of becoming infected every time it receives a packet from an infected neighbor. We also assume that infected nodes are recovered at a rate $\mu$, which we fix to 1 for most of the simulations. After a transient time, we compute the average density of infected individuals, $\rho$, which is the prevalence of disease in the system.

\begin{acknowledgments}
Y. M. is supported by MICINN (Spain) through the Ram\'on y Cajal
Program.  This work has been partially supported by MICINN through
Grants FIS2006-12781-C02-01, and FIS2008-01240. A.A. also acknowledges support from the Director, Office of Science, Computational and Technology Research, U.S. Department of Energy under Contract No. DE-AC02-05CH11231. S. M. thanks M. Meloni for useful discussions.
\end{acknowledgments}

\newpage

\begin{thebibliography}{99}

\bibitem{blmch06}
Boccaletti, S., Latora, V., Moreno, Y., Ch\'avez, M., Hwang,  D.~U. (2006) Complex Networks: Structure and Dynamics. {\em Phys. Rep.} 424:175-308.

\bibitem{dgm08}
Dorogovtsev, S.~N, Goltsev, A.~V., Mendes, J.~F.~F. (2008) Critical phenomena in complex networks. {\em Rev. Mod. Phys.} 80:1275-1336.

\bibitem{ba99}
Barab\'asi, A.~L., Albert, R. (1999) Emergence of scaling in random networks. {\em Science} 286: 509-512.

\bibitem{psv01}
Pastor-Satorras, R., Vespignani, A. (2001) Epidemic spreading in scale-free networks. {\em Phys. Rev. Lett.} 86:3200-3203.

\bibitem{llm01}
LLoyd, A.~L., May, R.~M. (2001) How viruses spread among computers and people. {\em Science} 292:1316-1317.

\bibitem{mpsv01}
Moreno, Y., Pastor-Satorras, R. Vespignani, A. (2002) Epidemic outbreaks in complex heterogeneous networks. {\em Eur. Phys. J. B} 26:521-529.

\bibitem{n02}
Newman, M.~E.~J. (2002) The spread of epidemic disease on networks. {\em Phys. Rev. E} 66: 016128.

\bibitem{bbpsv04}
Barth\'elemy, M., Barrat, A., Pastor-Satorras, R., Vespignani, A. (2004) Velocity and hierarchical spread of epidemic outbreaks in scale-free networks. {\em Phys. Rev. Lett.} 92:178701.

\bibitem{glmp08}
Garde\~nes, J.G., Latora, V., Moreno, Y., Profumo, E.  (2008) Spreading of sexually transmitted diseases in heterosexual populations. {\em Proc. Nat. Acad. Sci. USA} 105: 1399-1404.

\bibitem{asbs00}
Amaral, L.~A.~N. , Scala, A., Barth\' elemy, M., Stanley, H.~E. (2000) Classes of small-world networks. {\em Proc. Nat. Acad. Sci. USA} 97:11149-11152.

\bibitem{nfb02}
Newman, M.~E.~J., Forrest, S., Balthrop, J. (2002) Email networks and the spread of computer viruses. {\em Phys. Rev. E} 66:035101.

\bibitem{han_etal09} Han, J.-D. J., Bertin, N.,  Hao, T., Goldberg, D. S. et al (2004) Evidence for dynamically organized modularity in the yeast protein-protein interaction network. {\em Nature} 430: 88-93.

\bibitem{stanley}
Liljeros F., Edling C.~R., Amaral L.~A.~N., Stanley H.~E., Aberg Y. (2001) The Web of Human Sexual Contacts. {\em Nature} 411:907-908.

\bibitem{egamstw04}
Eubank, S., Guclu, H., Anil-Kumar, V.S., Marathe, M.V., Srinivasan, A., Toroczkai. Z., Wang, N.
(2004) Modelling disease outbreaks in realistic urban social networks. {\em Nature} 429:180Ð184.

\bibitem{cpv07} Colizza, V., Pastor-Satorras, R., Vespignani A. (2007) Reaction-diffusion processes and metapopulation models in heterogeneous networks. {\em Nature Physics} 3, 276-282.

\bibitem{cv08}
Colizza, V., Vespignani, A. (2008) Epidemic modeling in metapopulation systems with heterogeneous coupling pattern: Theory and simulations. {\em Journal of Theoretical Biology} 251, 450-467.

\bibitem{cbbv07} Colizza, V., Barrat, A., Barth\'elemy, M.,  Vespignani, A. (2007) Predictability and epidemic pathways in global outbreaks of infectious diseases: the SARS case study. {\em BMC Medicine} 5:34.

\bibitem{m07} Murray, J.~D. (2007) {\em Mathematical Biology} (Springer-Verlag, 3rd Edition).

\bibitem{skb08} Serrano, M.~A. , Krioukov, D., Bogu\~n\'a, M. (2008) Self-Similarity of Complex Networks and Hidden Metric Spaces. {\em Phys. Rev. Lett.} 100:078701.

\bibitem{bkc09}
Bogu\~n\'a, M.,  Krioukov, D., Claffy, K.~C. (2009) Navigability of Complex Networks. {\em Nature Physics} 5:74-80.

\bibitem{bk09}
Bogu\~n\'a, M.,  Krioukov, D. (2009) Navigating Ultrasmall Worlds in Ultrashort Time. {\em Phys. Rev. Lett.} 102:058701.

\bibitem{gdvca02}
Guimer\`a, R., D\'{\i}az-Guilera, A., Vega-Redondo, F., Cabrales, A., Arenas, A. (2002) Optimal network topologies for local search with congestion. {\em Phys. Rev. Lett.} 89:248701.

\bibitem{nr04} Noh, J. D.,  Rieger, H. (2004) Random Walks on Complex Networks. {\em Phys. Rev. Lett.} 92:118701.

\bibitem{mglm08} Meloni, S., Garde\~nes, J.G., Latora, V., Moreno, Y. (2008) Scaling Breakdown in Flow Fluctuations on Complex Networks. {\em Phys. Rev. Lett.} 100:208701.

\bibitem{zlpy05} Zhao,L., Lai, Y.-C., Park, K., Ye, N. (2005) Onset of traffic congestion in complex networks. {\em Phys. Rev. E} 71:026125.

\bibitem{sclts07} Sreenivasan, S., Cohen, R., Lopez, E., Toroczkai, Z., Stanley, H. E. (2007) Structural Bottlenecks for Communication in Networks. {\em Phys. Rev. E} 75:036105.

\bibitem{rmit08} Rosato, V., Meloni, S., Issacharoff, L., Tiriticco, F. (2008) Is the topology of the internet network really fit to its function? {\em Physica A} 387:1689-1704.

\bibitem{vespbook08}
Pastor-Satorras, R., Vespignani, A. (2004) {\em Evolution and Structure of the Internet: a statistical physics approach.} (Cambridge University Press).

\bibitem{rogerair}
Guimer\`{a}, R., Mossa, S., Turtschi, A. \& Amaral, L. A. N. (2005) The worldwide air transportation network: Anomalous centrality, community structure, and cities' global roles. {\em Proc. Natl. Acad. Sci. USA} 102:7794-7799.

\bibitem{gekg04}
Grais, R.~F.,  Ellis, J.~H., Kress, A.,  Glass, G.~E. (2004) Modeling the spread of annual influenza epidemics in the U.S.: The potential role of air travel. {\em Health Care Management Science} 7:127-134.

\end{thebibliography}
\end{document}